%% file: main.tex
\documentclass[sigconf]{acmart}
\acmConference[ESEC/FSE 2022]{The 30th ACM Joint European Software Engineering Conference and Symposium on the Foundations of Software Engineering}{14 - 18 November, 2022}{Singapore}
\AtBeginDocument{
  \providecommand\BibTeX{{
    \normalfont B\kern-0.5em{\scshape i\kern-0.25em b}\kern-0.8em\TeX}}}

\input{macro}

\begin{document}

\title{Less Training, More Repairing Please: \\Revisiting Automated Program Repair via Zero-shot Learning}

\author{Chunqiu Steven Xia}
\affiliation{
  \institution{University of Illinois Urbana-Champaign}
  \country{}
}
\email{chunqiu2@illinois.edu}

\author{Lingming Zhang}
\affiliation{
  \institution{University of Illinois Urbana-Champaign}
  \country{}
}
\email{lingming@illinois.edu}

\begin{abstract}

Due to the promising future of \aprfull (\apr), researchers have proposed various \apr techniques, including \heuristic, \template, and \constraint techniques. Among such classic \apr techniques, \template techniques have been widely recognized as state of the art. However, such \template techniques require predefined templates to perform repair, and their effectiveness is thus limited. To this end, researchers have leveraged the recent advances in Deep Learning to further improve \apr. Such \learning techniques typically view \apr as a \nmtfull problem, using the buggy/fixed code snippets as the source/target languages for translation. In this way, such techniques heavily rely on large numbers of high-quality bug-fixing commits, which can be extremely costly/challenging to construct and may limit their edit variety and context representation.%

In this paper, we aim to revisit the \learning \apr problem, and propose \tech, the first \emph{\cloze} \apr approach to directly leveraging large pre-trained code models for \apr without any fine-tuning/retraining on historical bug fixes. \emph{Our main insight is instead of modeling what a repair edit should look like (i.e., a NMT task), we can directly predict what the correct code is based on the context information (i.e., a cloze task).} Although our approach is general and can be built on various pre-trained code models, we have implemented \tech as a practical multilingual \apr tool based on the recent \codebert model. Our evaluation of \tech on the widely used \dfj benchmark \emph{shows for the first time that \learning \apr without any history bug fixes can substantially outperform state-of-the-art \apr techniques}. We also studied the impact of different design choices and show that \tech performs even better on a newer version of \dfj (2.0) with 3.3X more fixes than best performing baseline, indicating that \tech can potentially avoid the dataset-overfitting issue of existing techniques. Additionally, we demonstrate the multilingual repair ability of \tech by evaluating on the \quix dataset where \tech achieved the state-of-the-art results on both Java and Python versions.

\end{abstract}

\maketitle

\input{intro}

\input{background}

\input{approach}
\input{experiment}

\input{result}
\input{threats}
\input{conclude}

\balance
\bibliographystyle{ACM-Reference-Format}
\bibliography{reference, lingming}

\end{document}

%% file: macro.tex
\usepackage{xspace}
\usepackage{makecell}
\newcommand{\CodeIn}[1]{{\small \texttt{#1}}}

\newcommand{\mypara}[1]{\vspace{.03in}\noindent \textbf{#1}}
\newcommand{\Comment}[1]{}
\newcommand{\tech}{AlphaRepair\xspace} %
\newcommand{\cloze}{cloze-style\xspace}

\newcommand{\codebert}{CodeBERT\xspace}

\newcommand{\inputseqfull}{X = \{x_1, x_2, ... x_n\}\xspace}

\newcommand{\inputseqone}{x_i\xspace}
\newcommand{\maskseq}{X_{masked}\xspace}
\newcommand{\maskcode}{C_{masked}\xspace}
\newcommand{\maskword}{W_{masked}\xspace}
\newcommand{\predictor}{P\xspace}
\newcommand{\lossmlm}{\mathcal{L}_{\mlm}\xspace}
\newcommand{\lossmlmbi}{\mathcal{L}_{bimodal\_\mlm}\xspace}
\newcommand{\blinelength}{L\xspace}

\newcommand{\tjs}{temp joint score\xspace}
\newcommand{\js}{joint score\xspace}
\newcommand{\beamwidth}{N\xspace}
\newcommand{\tokenslength}{n\xspace}
\newcommand{\tokensgeneratedlength}{p\xspace}
\newcommand{\codebertpredictor}{C\xspace}
\newcommand{\codebertpredictortemp}{C^*\xspace}
\newcommand{\generatedseqfull}{T = \{t_1, ... t_p, MASK_{p+1}, ... MASK_{n}\}\xspace}
\newcommand{\generatedseq}{T}
\newcommand{\generatedseqone}{t}
\newcommand{\generatedseqi}{t_i}
\newcommand{\generatedseqp}{t_p}
\newcommand{\generatedseqprev}{\{t_1, ... t_{p-1}\}\xspace}
\newcommand{\generatedseqoneless}{\{t_1, ... t_{p-1}, t_p, MASK_{p+1}, ... MASK_{n}\}\xspace}
\newcommand{\generatedseqafter}{\{MASK_{p+1}, ... MASK_{n}\}\space}
\newcommand{\fullgeneratedseq}{T = \{t_1, t_2, ... t_n\}\xspace}
\newcommand{\recoder}{Recoder\xspace}
\newcommand{\deepdebug}{DeepDebug\xspace}
\newcommand{\cure}{CURE\xspace}
\newcommand{\coconut}{CoCoNuT\xspace}
\newcommand{\dlfix}{DLFix\xspace}
\newcommand{\sequencer}{SequenceR\xspace}
\newcommand{\tbar}{TBar\xspace}
\newcommand{\prapr}{PraPR\xspace}
\newcommand{\avatar}{AVATAR\xspace}
\newcommand{\simfix}{SimFix\xspace}
\newcommand{\fixminer}{FixMiner\xspace}
\newcommand{\capgen}{CapGen\xspace}
\newcommand{\jaid}{JAID\xspace}
\newcommand{\sketchfix}{SketchFix\xspace}
\newcommand{\nopol}{NOPOL\xspace}
\newcommand{\jgenprog}{jGenProg\xspace}
\newcommand{\jmutrepair}{jMutRepair\xspace}
\newcommand{\jkali}{jKali\xspace}

\newcommand{\apr}{APR\xspace}
\newcommand{\aprfull}{Automated Program Repair\xspace}
\newcommand{\nmt}{NMT\xspace}
\newcommand{\nmtfull}{Neural Machine Translation\xspace}
\newcommand{\nlp}{NLP\xspace}
\newcommand{\nlpfull}{Natural Language Processing\xspace}
\newcommand{\mlm}{MLM\xspace}
\newcommand{\mlmfull}{Masked Language Model\xspace}

\newcommand{\dfj}{Defects4J\xspace}
\newcommand{\quix}{QuixBugs\xspace}

\newcommand{\learning}{learning-based\xspace}
\newcommand{\Learning}{Learning-based\xspace} %
\newcommand{\template}{template-based\xspace}
\newcommand{\heuristic}{heuristic-based\xspace}
\newcommand{\constraint}{constraint-based\xspace}
\newcommand{\maskline}{mask line\xspace}
\newcommand{\Maskline}{Mask line\xspace}
\newcommand{\masktoken}{mask token\xspace}

\newcommand{\distance}{5pt}

%% file: intro.tex
\section{Introduction}

Software systems are all-pervasive in everyday life from monitoring financial transactions~\cite{finance}, controlling transportation systems~\cite{webster1993} to aiding healthcare tools~\cite{healthcare}. Software bugs in these systems can affect people around the globe and cost billions in financial losses~\cite{buglossarticle}. 
To fix such bugs, developers often need to invest significant manual effort, e.g., it is estimated that developers spend 35 to 50\% of their time debugging software systems~\cite{debuggingtime}. To reduce manual debugging efforts, \aprfull (\apr) techniques have been proposed to automatically generate patches to fix the bugs~\cite{gazzola2019aprsurvey}.

A popular approach for \apr is Generate and Validate (G\&V)~\cite{qi2015gv, ghanbari2019prapr, lou2020can, le2016hdrepair, legoues2012genprog, wen2018capgen, hua2018sketchfix, martinez2016astor, koyuncu2020fixminder, liu2019tbar, liu2019avatar}. To start off, fault localization~\cite{wong2016fl, abreu2007ochiai, zhang2013injecting, papadakis2015metallaxis, li2019deepfl, li2017transforming} is often used to reduce the search space by computing the suspicious program locations that likely caused the bug. Using these potential buggy locations, the G\&V techniques will generate a list of candidate patches. Each candidate patch is compiled and validated against the test suite. Patches that successfully pass all tests are called \emph{plausible patches}. However, tests often cannot cover all possible behaviors of the program~\cite{qi2015gv}, hence a plausible patch might still fail under other inputs. Therefore, plausible patches are further inspected by developers to determine the final \emph{correct patches} that correctly fix the underlying bug. 

Depending on how patches are generated, traditional \apr techniques can be categorized into \heuristic~\cite{legoues2012genprog, le2016hdrepair, wen2018capgen}, \constraint~\cite{mechtaev2016angelix, le2017s3, demacro2014nopol}, and \template~\cite{hua2018sketchfix, martinez2016astor, koyuncu2020fixminder, liu2019tbar, liu2019avatar}. Among all traditional techniques, \template \apr techniques, which leverage pre-defined fix patterns to transform buggy code snippets into correct ones, have been widely recognized as state of the art~\cite{ghanbari2019prapr, liu2019tbar, benton2020effectiveness}.
These fix patterns target specific types of bugs (e.g., null pointer exception) and patterns in the source code, and are often crafted by human experts. While effective, there is an upper limit to the number of candidate patches that such pre-defined templates can generate. Therefore, to increase the expressiveness of the edits, researchers have recently utilized Machine Learning (ML) or Deep Learning (DL) techniques for patch generation~\cite{li2020dlfix, chen2018sequencer, jiang2021cure, lutellier2020coconut, zhu2021recoder}.

\Learning \apr techniques often leverage the recent advances in DL to train a neural network that transforms the original buggy program into the fixed version. These techniques~\cite{chen2018sequencer, li2020dlfix, lutellier2020coconut, jiang2021cure, zhu2021recoder, drain2021deepdebug} typically view the problem of program repair as a \nmtfull (\nmt)~\cite{sutskever2014sequence} problem and use \nmt models from the field of \nlpfull (\nlp), where the model input is a language sequence and the output is a translated sequence in another language. Researchers have used \nmt models for program repair where instead of translating natural languages, the models aim to turn a buggy code into the fixed version. These \nmt models are typically made up of an \emph{encoder} and \emph{decoder} pair where the encoder captures the buggy code elements with its context and the decoder takes in the encoded representation and generates a fixed version. To facilitate \apr, such models \emph{must} be trained using pairs of buggy and patched code. Despite being an effective APR approach, existing \learning techniques face the following issues:

\emph{1) Quality of training data.} Current \learning \apr tools require training or fine-tuning of the models by using historical bug fixes, i.e., pairs of buggy and patch code. This data is usually obtained by scraping open-source projects to find specific commits that are about bug fixes. However, this relies on various handcrafted heuristics. For example, to find the bug fixing commits, keywords such as \textit{bug, fix, patch, solve, issue, problem} are often used to filter the commits~\cite{zhu2021recoder, lutellier2020coconut, dallmeier2007benchmark, jiang2019infer}. Individual bug-fixing commits can also include unrelated edits such as refactoring or new feature implementation~\cite{jiang2021extract}. As a result, the extracted data can contain various irrelevant commits and unrelated code changes within bug fixing commits, adding noise to the training dataset. 

\emph{2) Quantity of training data.} Compared to large amount of open-source code snippets that are available in the wild, the amount of bug fixes is limited. To reduce the effect of the aforementioned issue of a commit containing irrelevant changes from bug fixes, \learning \apr tools usually limit the commits in their dataset to ones with few lines of changes~\cite{zhu2021recoder, li2020dlfix, lutellier2020coconut, jiang2021cure}, further limiting the amount of training data. By training on such limited historical fixes, current \learning tools might restrict the edit variety of their approach only on what is in their training data.

\emph{3) Context representation.} To provide a correct fix to a buggy code snippet, the context before and after are crucial in providing useful syntactic/semantic information. Current \learning \apr tools first pass the context including the buggy code elements into an encoder as plain texts~\cite{lutellier2020coconut, jiang2021cure} or structured representations~\cite{li2020dlfix, zhu2021recoder}. The encoded context is then used directly or combined with a separate encoding of the buggy code snippet as input to the decoder. However, this process is \textit{unnatural} since it is challenging for the models to distinguish the patch location within the context, or effectively merge the separate bug/context encodings. As a result, such techniques may miss intricate relations between a patch and its context, such as the proximity of each code element that provides important syntax/semantic information. 

\mypara{Our Work.} We present \tech{} -- the first \emph{\cloze} \apr approach that uses large pre-trained code models under a \emph{zero-shot learning} setting~\cite{lampert2009zsl} to directly generate patches, i.e., without any additional training or fine-tuning on bug-fixing datasets.
Different from all existing \learning \apr techniques, our main insight is that \emph{instead of modeling what a repair edit should look like, we can directly model/predict what the correct code is based on the context information} (like a \emph{cloze}~\cite{taylor1953cloze, clozewiki} or ``fill-in-the-blank'' task). In this way, our \cloze \apr can avoid all above limitations of the existing techniques: 1) it completely frees \apr from historical bug fixes, 2) it can simply get trained on all possible open-source projects in the wild for massive training, 3) it is directly pre-trained to model patches based on the surrounding context, and thus can effectively encode the intricate relations between patches and their contexts. Furthermore, while it is non-trivial to adapt prior APR techniques for a
new language (due to a huge amount of code/data engineering work for preparing historical
bug fixes), under our \cloze \apr, extending APR to a new language can be as simple as
mining a new code corpus in the new language!

While our \cloze \apr is generalizable to various pre-trained models, in this paper, we implement \tech with one recent pre-trained code model, \codebert~\cite{feng2020codebert}. Unlike current \learning \apr tools which use limited numbers of bug fixes as training data, \codebert is directly pre-trained using millions of code snippets from open-source projects, allowing it to provide a variety of edit types to fix different bugs. We directly leverage the initial training objective of \mlmfull (\mlm)~\cite{devlin2019bert} - predicting/recovering randomly masked out tokens used in \codebert to perform zero-shot learning for \apr.
We first prepare model inputs where each potentially buggy line is replaced with a \maskline. We then query \codebert to fill the \maskline with replacement tokens to produce candidate patches for a buggy program input. This allows us to directly perform zero-shot learning (with no fine-tuning) since we use the same tasks as defined in \mlm -- instead of predicting random \masktoken{s}, we generate predictions by masking out only the buggy code snippet. Note that to improve the patch search space, the \maskline{s} are designed to systematically reuse parts of the buggy line. Furthermore, the bidirectional nature of \codebert (and \mlm) also allows us to naturally capture both the contexts before and after the buggy location for effective patch generation.  

\mypara{Contribution.} This paper makes the following contributions:%
\begin{itemize}
    \item \textbf{Direction/Dimension} This paper opens a new dimension for \cloze \apr to directly query large pre-trained code models under a zero-shot learning setting. Compared with existing \learning \apr techniques, our approach does not need any additional fine-tuning/retraining using historical bug fixes and can be easily adopted for multiple programming languages. Additionally, we demonstrate the efficacy of directly applying large pre-trained models for generating code fixes for real-world systems, which previously were mainly tested on generating simple/small programs~\cite{austin2021program, chen2021codex}.  
    \item \textbf{Technique} While our idea is general and can leverage various existing pre-trained code models, we have built \tech as a practical \apr tool based on the recent \codebert model. We leveraged the original training objective of \codebert using specific inputs of \maskline{s} for direct program repair. We used a variety of different \maskline templates and further propose probabilistic patch ranking to boost \apr.
    \item \textbf{Extensive Study} We have compared \tech with state-of-the-art Java \apr tools (both traditional and \learning) on \dfj~\cite{just2014dfj}. The results show that \tech can outperform all existing tools on the widely studied \dfj 1.2, improving the number of fixed bugs from 68 to 74 and fixing 8 unique bugs that no prior work can fix. More surprisingly, \tech even fixes 3.3X more bugs than the best baseline on the newly included bugs in \dfj 2.0, demonstrating that \tech can avoid the dataset-overfitting issue of existing \apr techniques. Moreover, we have also studied \tech on both the Java and Python versions of the widely studied \quix dataset~\cite{lin2017quixbug}. The results not only confirm that \tech can outperform all existing \apr techniques (in both Java and Python), but also demonstrate the multilingual capability of \tech.   
\end{itemize}

%% file: background.tex
\section{Background}

\begin{figure}
    \includegraphics[width=0.95\linewidth]{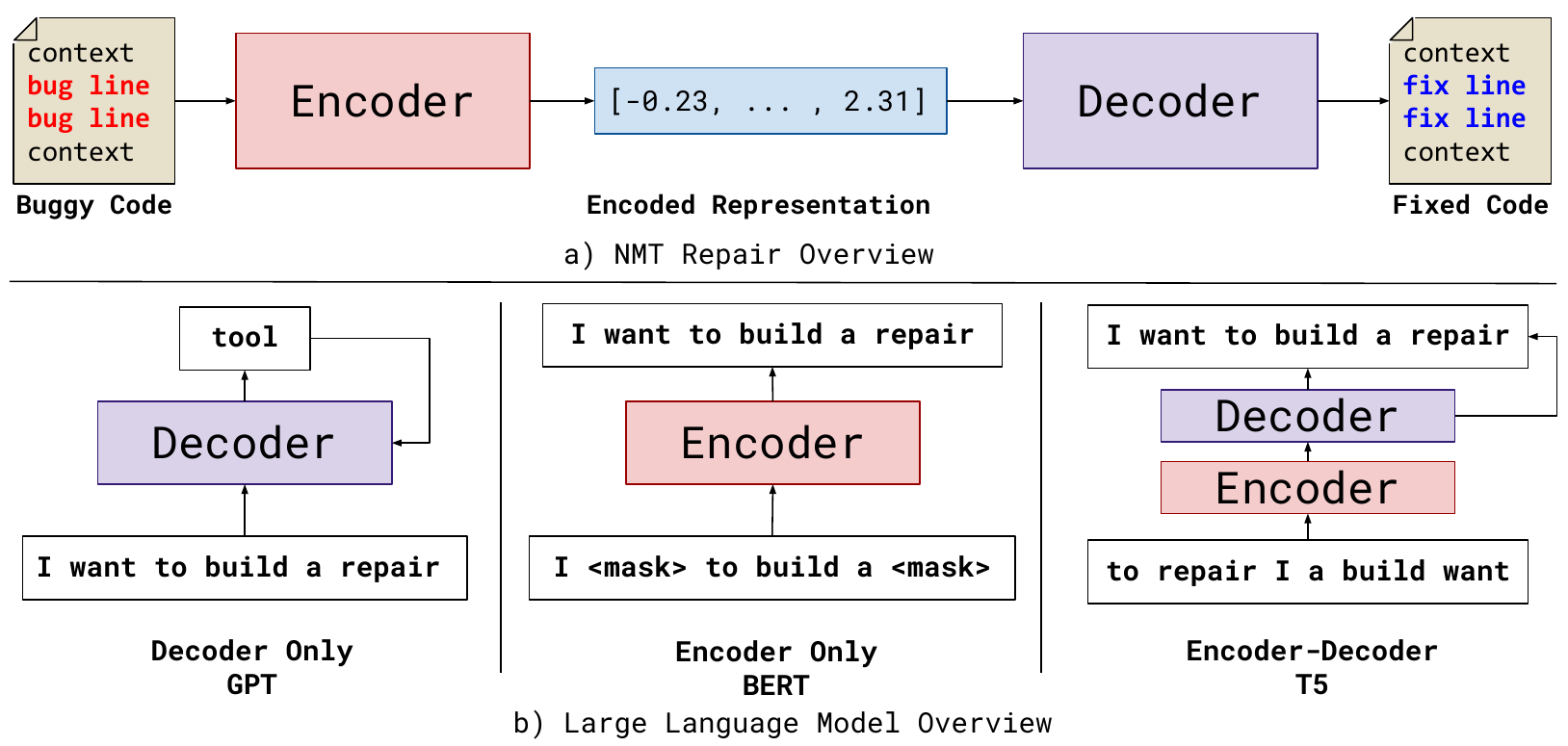}
    \caption{\nmt and large language model overview}
    \label{fig:background}
\end{figure}

\subsection{\Learning \apr}

Deep Learning (DL)~\cite{GoodBengCour16DL} is a powerful learning mechanism to learn from large amounts of data used in many domains of Machine Learning. Researchers have leveraged DL techniques for \apr by viewing the problem of program repair as a \nmtfull (\nmt)~\cite{sutskever2014sequence} task of turning a buggy code snippet into a patched version. These \learning \apr tools are typically built using the encoder-decoder architecture. Figure~\ref{fig:background}a shows the workflow of these \learning repair tools. The encoder takes in the buggy line and its surrounding context as input and outputs an encoded representation. The decoder uses the encoded representation to generate a new code line to replace the buggy line. \sequencer~\cite{chen2018sequencer} is a sequence-to-sequence network built using a Long Short-Term Memory (LSTM)~\cite{hochreiter1997lstm} encoder and decoder network for program repair. \dlfix~\cite{li2020dlfix} encodes the input code as an Abstract Syntax Tree then uses a tree-based Recurrent Neural Network~\cite{rumelhart1985rnn} as a part of the encoder and decoder network to generate patches. \coconut~\cite{lutellier2020coconut} proposes a new \nmt architecture which encodes the context and buggy line separately. It also leverages multi-stage attention to increase the amount of important information passed on from the encoder to the decoder. In order to improve the syntactic correctness, before training on code fix pairs, \cure~\cite{jiang2021cure} first pre-trains the \nmt model on a large corpus of developer code. Furthermore, \cure uses a static checking strategy to only generate patches with valid identifiers. More recently, \recoder~\cite{zhu2021recoder} modifies the \nmt model by using a syntax-guided decoder designed to generate syntactically correct edits on an input buggy code snippet.

Recent study has shown that \learning \apr tools can achieve the state of the art in \apr~\cite{zhu2021recoder, jiang2021cure}. However, they are still limited by their need for pairs of buggy and fixed versions as training data. These buggy and fixed code pairs are challenging to obtain as handcrafted heuristics are used to identify bug fixing commits and individual commits can contain other code changes apart from bug fixing, adding noise to the dataset. Additionally, \learning \apr tools learn bug patterns and corresponding patch fixes from the training data in order to automatically generate patches for unseen buggy code. This means it is hard for \learning \apr tools to generalize to fix patterns that are not present in their training dataset. Furthermore, it can be tricky for current \learning \apr tools to encode the context surrounding buggy code elements causing the generated patches to miss important syntax and semantic information. Interestingly, although pre-trained code models have also been adopted for \apr recently~\cite{jiang2021cure, drain2021deepdebug, mashhadi2021CodeBertRepair}, they are still leveraged to learn from a large number of historical bug fixes, thus still suffering from the above limitations of existing \learning techniques.

In this paper, we address these issues by using large pre-trained code model, trained on massive amount of open-source code, directly for \cloze \apr without the need to train or fine-tune on any smaller dataset of buggy and fixed code. 

\subsection{Large Pre-trained Code Models}

Recent popularity in \nlpfull (\nlp) has led to development of large pre-trained models that use large amounts of data. To make use of the massive unlabeled training data, pre-trained models apply self-supervised objectives, e.g., \textbf{\mlmfull (\mlm)} - where some training data is artificially masked and the training objective is to predict/recover the real data. A common component of large pre-trained language models is a Transformer~\cite{vaswani2017attention}. It contains an encoder made up of multiple differentiable self-attention layers in order to learn representation and also a decoder used to generate output. Figure~\ref{fig:background}b shows the three categories of large pre-trained language models. GPT~\cite{radford2019gpt2} is a large generative model which uses only the \emph{decoder} component to predict the next token output given all previous tokens. This type of decoder is autoregressive where a sequence is generated by iteratively inputting all previous tokens in order to generate the next one. BERT~\cite{devlin2019bert} is another type of large pre-trained model which contains only the \emph{encoder} component. BERT is designed to learn a representation of the data and is trained using the \mlm objective. A small percentage of tokens in the training data will be replaced by a \masktoken, where the goal is to train BERT to predict the true value of the \masktoken. To combine the usage of both encoder and decoder, \emph{encoder-decoder}
models have also been used to build large pre-trained language models. Models such as T5~\cite{raffel2020exploring} are designed for sequence-to-sequence tasks where the training objective aims to recover the correct output sequence given the original input (English to French, corrupted to uncorrupted, etc). These large pre-trained language models can be fine-tuned for downstream \nlp tasks such as text summarization~\cite{liu2019finetune}, text classification~\cite{yang2020xlnet}, as well as question and response text generation~\cite{clark2019boolq}.

Researchers have extended encoder, decoder
and encoder-decoder models to build large pre-trained models for various \emph{programming language} tasks.
CodeGPT~\cite{lu2021codexglue} adopts the original GPT architecture and trains a generative model from scratch using Python and Java functions.
Codex~\cite{chen2021codex} is a GPT-based code model created by fine-tuning a larger GPT-3 model~\cite{brown2020gpt3} for generating Python functions based on natural language descriptions. \codebert~\cite{feng2020codebert} and GraphCodeBERT~\cite{guo2021graphcodebert} are BERT-based models for programming tasks, and are trained using the \mlm training objective. GraphCodeBERT additionally encodes simple data flow information to aid in code representation. CodeTrans~\cite{elnaggar2021codetrans}, CodeT5~\cite{wang2021codet5} and PLBART~\cite{ahmad2021PLBART} are unified encoder-decoder models~\cite{raffel2020exploring} which uses denoising sequence-to-sequence training objectives to pre-train both the encoder and decoder for various coding tasks.%

In this paper, we directly use large pre-trained models for \cloze \apr via zero-shot learning. While our \cloze \apr idea is general and can be achieved using all above pre-trained models, we demonstrate its potential by using the simple \codebert model. \codebert is trained using the \mlm objective which can be used to generate replacement code for buggy code snippets. Also, \codebert is bidirectional in nature, allowing it to capture both contexts before and after for patch generation.

%% file: approach.tex
\section{Approach}

\begin{figure*}
    \includegraphics[width=0.8\linewidth]{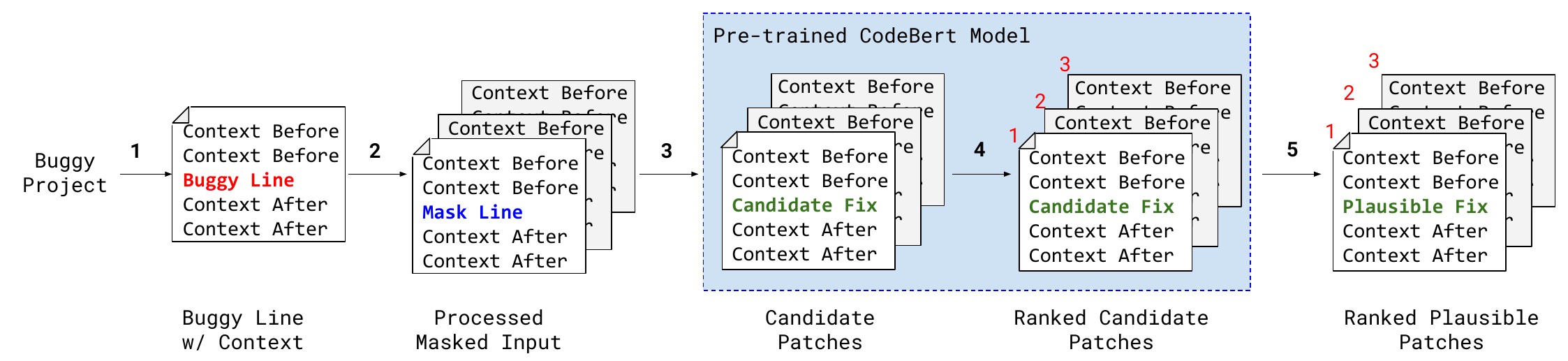}
    \caption{\tech overview}
    \label{fig:overview}
\end{figure*}

In this section, we introduce \emph{\cloze} \apr,  a new direction for \learning \apr that directly learns from the large number of code snippets available in the wild to generate patches. 
Different from all prior \learning \apr techniques, instead of viewing \apr as a task of turning a buggy code snippet into a fixed code snippet, we can directly learn what the \emph{correct code} should look like given its surrounding context. We view the problem as a \emph{cloze task}~\cite{clozewiki, taylor1953cloze} where we replace the buggy snippet with blanks/masks and query pre-trained models~\cite{baevski2019cpretraining} to fill the blanks with the correct code snippet.
\emph{This problem setup does not require access to any dataset containing pairs of buggy and patched code versions and our approach can directly make use of the existing large pre-trained models to automatically generate code repairs.}

Although our approach is general, in this work, we re-purpose the recent \codebert \cite{feng2020codebert} model for program repair. The training objective for \codebert uses \mlmfull (\mlm)~\cite{devlin2019bert} where given an input sequence of tokens $\inputseqfull$, a random set of tokens in $X$ is replaced with \emph{\masktoken{s}} to generate a new mask sequence $\maskseq$. The training goal is to output the original tokens in $\maskseq$ which have been masked out, i.e. recover $X$ given $\maskseq$. Given predictor $\predictor$ which outputs the probability of a token, the \mlm loss function can be formulated as:
\begin{equation}
    \label{eq:mlmloss}
    \lossmlm = \sum_{i \in masked}-\log(\predictor(\inputseqone | \maskseq))
\end{equation}

The \mlm training objective allows us to directly use \codebert for program repair where instead of randomly masking out tokens, we mask out all tokens which are part of the buggy code snippet. We then use \codebert under a zero-shot learning setting for program repair where we recover the correct tokens in place of the mask buggy tokens. As a result, \tech does not require any additional retraining or fine-tuning stage on bug fixing datasets since the \mlm training is done as part of the pre-training tasks in \codebert.
While our basic idea is applicable for \apr at different levels, following state-of-the-art \learning \apr tools~\cite{jiang2021cure, lutellier2020coconut, zhu2021recoder}, we focus on single line patches in this work. Figure \ref{fig:overview} provides an overview of our approach:
\begin{itemize}
    \item \textbf{Step 1 (Section \ref{input_processing})}: We first take in a buggy project and separate the surrounding context and the buggy line according to fault localization information. We encode both the context before and after into token representations. Additionally, we also encode the buggy line as a comment in the natural language input for \codebert.
    \item \textbf{Step 2 (Section \ref{mask_generation})}: Using the buggy line, we generate multiple \textbf{\maskline{s}} using templates (replace entire line, replace starting/ending part of line, etc). Each \maskline replaces the buggy line and is tokenized together with the surrounding context as inputs to \codebert. 
    \item \textbf{Step 3 (Section \ref{patch_generation})}: We iteratively query \codebert to generate candidate patches using \maskline{s}. Each patch replaces the \maskline with a generated code line. 
    \item \textbf{Step 4 (Section \ref{patch_reranking})}: We use \codebert again to provide patch ranking by computing the score of the generated patch using the joint probability of the generated tokens. 
    \item \textbf{Step 5 (Section \ref{patch_validation})}: We compile each candidate patch and validate it against the test suite. Finally, we output a list of plausible patches for developers to examine. 
\end{itemize}

\subsection{Input Processing}
\label{input_processing}

\begin{figure}[t!]
    \includegraphics[width=\linewidth]{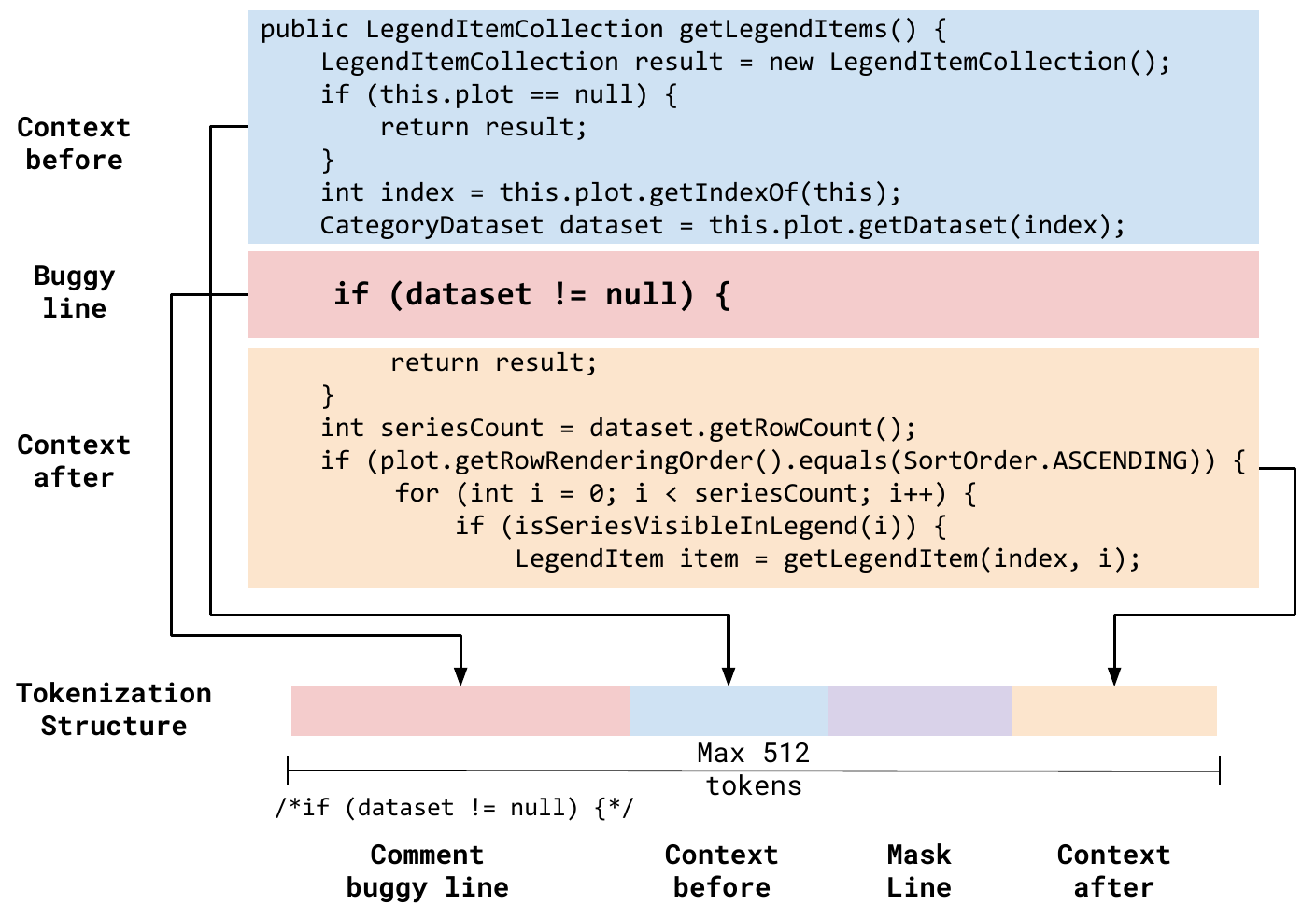}
    \caption{Example input for \tech}
    \label{fig:input_processing_example}
\end{figure}

To generate the inputs for \tech, first we extract the buggy line and surrounding context from the buggy project. We use the \codebert tokenizer which is built using byte-level byte pair encoding (BBPE) -- a technique to reduce the size of the vocabulary by breaking uncommon long words into subwords that are found commonly in the corpus \cite{wang2019bbpe}. BBPE has been used in various models and shown to mitigate Out of Vocabulary issues \cite{liu2019roberta, radford2019gpt2}. 

Figure \ref{fig:input_processing_example} provides an example input of a buggy program. We first define our tokenization structure, a list of tokens as inputs for \codebert. We tokenize both the context before and after and sandwich the \maskline (\emph{placeholders} for what \codebert will predict, described in Section \ref{mask_generation}) between them. For program repair, the buggy line itself is also important to generate a patch. However, it does not make sense to include it as a part of the context since we aim to generate code to replace it. 

To capture the encoding for the buggy line, we make use of the bimodal nature of \codebert where it can take in both programming language and also natural language (comments). We transform the original buggy line into a comment by surrounding it with the block comment characters (\CodeIn{/* comment */}). Recall Equation \ref{eq:mlmloss} which describes the basic \mlm loss function, where $\maskseq$ is a mask sequence. $\maskseq$ in \codebert concatenates both natural language and code tokens such that $\maskseq =  \maskword, \maskcode$, where $\maskword$ and $\maskcode$ are mask sequences of natural language and code token sequences. The original \mlm loss function is now:
\begin{equation}
    \label{eq:bi_mlmloss}
    \lossmlmbi = \sum_{i \in masked}-\log(\predictor(\inputseqone | \maskword, \maskcode))
\end{equation}

This way \codebert learns both modalities of function representation (natural language and function code). \tech makes use of this additional understanding and transforms the buggy line into a comment. Together the comment buggy line, context before, \maskline and context after are tokenized as input for \codebert. To maximize the context we can encode, we start from the buggy line and increase the context size (lines away from the buggy code) until we reach the maximum \codebert input token size of 512.

\subsection{Mask Generation}
\label{mask_generation}

\begin{figure}
    \includegraphics[width=\linewidth]{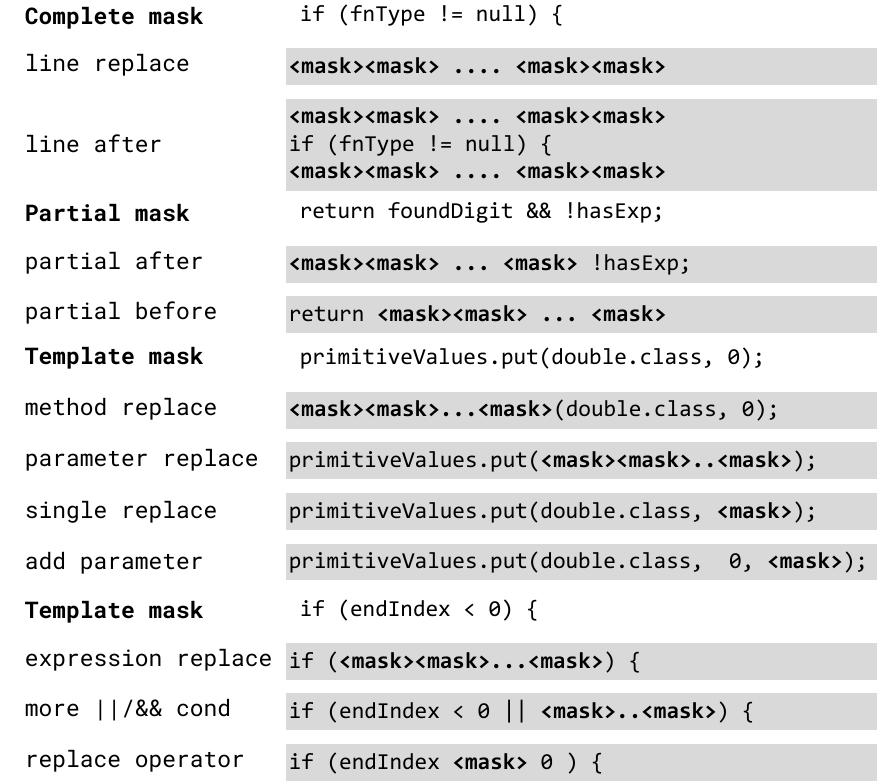}
    \caption{Different strategies to generate \maskline{s}}
    \label{fig:mask_generation}
\end{figure}

In order to generate patches, we replace the buggy line with a \textbf{\maskline}. \Maskline is defined as a line with one or more \masktoken{s} - \CodeIn{<mask>}. We use \codebert to fill each \masktoken with a replacement code token and together the filled \maskline becomes a generated patch line. Figure \ref{fig:mask_generation} shows the 3 strategies we use to generate a \maskline: complete, partial and template mask. 

\mypara{Complete mask.} The simplest strategy is to replace the entire buggy line with a line containing only \masktoken{s}. We refer to this as line replacement since we ask \codebert to generate a new line to replace the buggy line. We also generate \maskline{s} where we add \masktoken{s} before/after the buggy line. These represent bug fixes where a new line is inserted before/after the buggy location. 

\mypara{Partial mask.} Another strategy of generating \maskline{s} is by reusing partial code from the buggy line. We first separate the buggy line into its individual tokens and then keep the last/first token and replace all other tokens before/after with the \masktoken{s}. We then repeat this process but append more tokens from the original buggy line to generate all the possible versions of partial \maskline{s}. 

For example, in the partial before strategy in Figure \ref{fig:mask_generation}, we start with generating a \maskline of \CodeIn{return <mask><mask>...<mask>} by keeping the first token of \CodeIn{return}. Then we generate another \maskline by keeping the first two tokens (\CodeIn{return foundDigit}) to generate \CodeIn{return foundDigit <mask><mask>...<mask>}. In total, we generate ($\blinelength - 1$)
number of \maskline{s}, where $\blinelength$ is the number of tokens in the buggy line, for both partial after and before generation method. 

This approach is motivated by patches where the fix will reuse parts of the buggy line. By prepending and appending the \maskline with a part of the buggy line, we can reduce the number of tokens \codebert needs to generate for a correct fix. Furthermore, the partial buggy code acts like initial starting point for \codebert to start generating tokens by providing important context. 

\mypara{Template mask.}
We implemented several \template \maskline generation strategies targeting conditional and method invocation statements as they are two of the most common bug patterns~\cite{pan2009pattern, ghanbari2019prapr, le2016hdrepair}. Additionally, several traditional \apr tools~\cite{demacro2014nopol, xiong2017acs, long2015spr, durieux2016dynamoth} focus solely on fixing conditional statement related bugs, showing the importance of targeting common bug patterns.
Unlike the previous 2 strategies, template mask can only be generated for specific buggy lines. The first set of templates are designed to target buggy method invocations. Method replacement will replace the method call name with \masktoken{s}. This represents asking \codebert to generate a replacement method call using the same parameters as before. We also use several parameter based changes:  replacing the entire inputs with \masktoken{s}, replacing one parameter with \masktoken{s}, and adding additional parameter(s) (more than one parameters can be added since we vary the number of \masktoken{s}, therefore \codebert can add multiple parameters).

We also designed template \maskline{s} for conditional statements in the form of a Boolean expression. We generate \maskline{s} that replace the entire Boolean expression or add additional and/or expressions by appending the statement with \masktoken{s}. Additionally, we also identify common operators (\CodeIn{<, >, >=, ==, \&\&, ||, etc}) and replace them directly in the buggy line with \masktoken{s}.

These \template \maskline generations are inspired by common fixes for many bugs~\cite{pan2009pattern, ghanbari2019prapr, le2016hdrepair} and also previous APR tools that utilize preset templates to fix bugs~\cite{hua2018sketchfix, martinez2016astor, koyuncu2020fixminder, liu2019tbar, liu2019avatar}. These simple generated templates serve a similar functionality to the partial masks in providing more starting code for \codebert to generate potential patches. For a fix that only needs to modify a small part of the buggy code, \codebert only needs to generate a small number of tokens using \maskline{s} from the template masks. By including a larger portion of the buggy code, we reduce the search space \codebert has to consider.

For each generated \maskline, we increase the number of generated tokens from 1 until the total number of tokens in that line becomes ($\blinelength + 10$) where $\blinelength$ is the number of tokens in the original buggy line. For example, if we use the ``partial-after'' strategy on a buggy line with $\blinelength$ of 12 and we keep the first 5 original tokens, we will vary the number of masked tokens from 1 to 17. This process is done for each masking strategy that we apply (except for the replace operator strategy where we only replace common operators with a single mask token). We apply the input processing step (Section \ref{input_processing}) for each \maskline to obtain a list of processed inputs for \codebert.

\subsection{Patch Generation}
\label{patch_generation}

\begin{figure}
    \includegraphics[width=\linewidth]{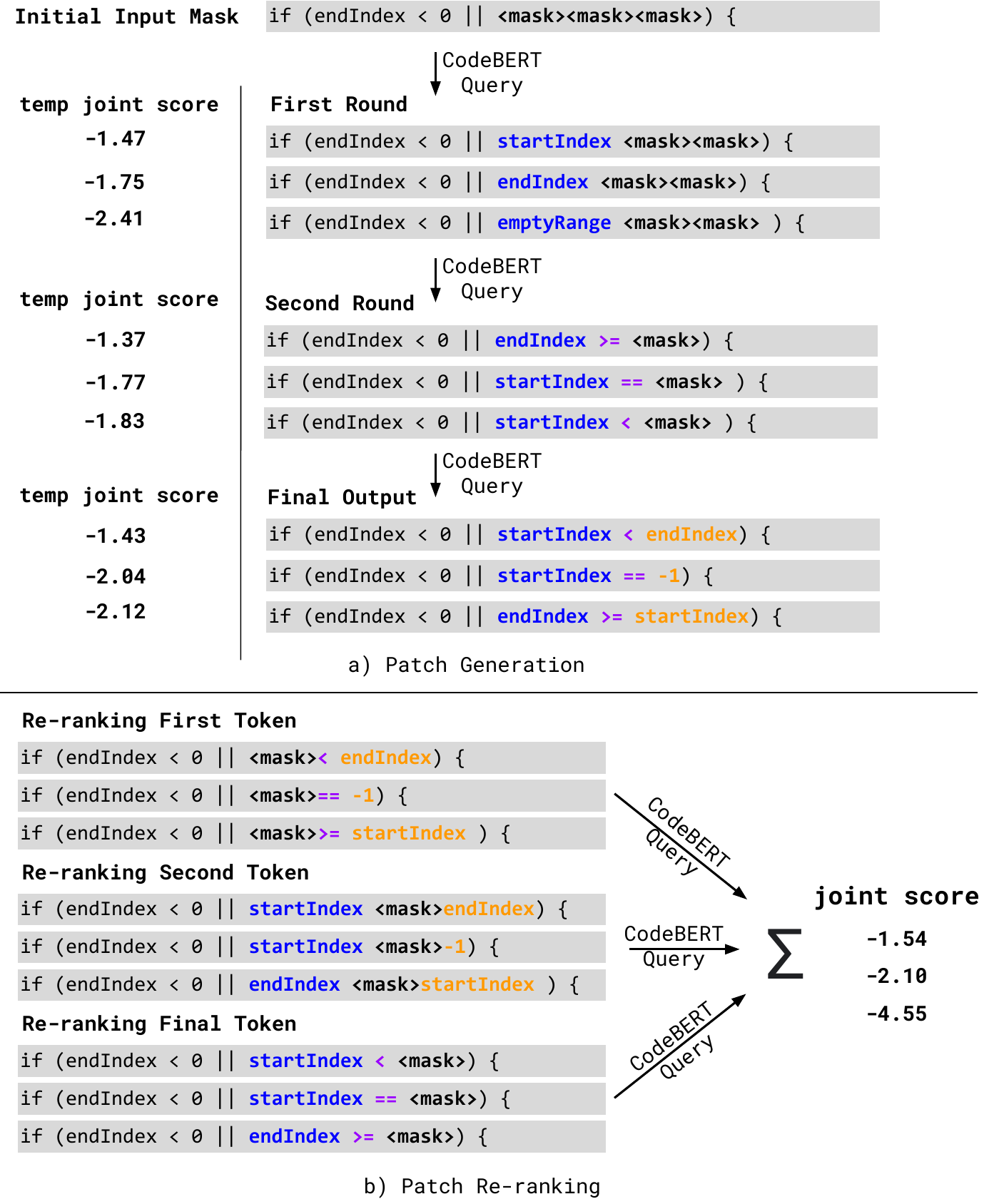}
    \caption{Patch generation and re-ranking example}
    \label{fig:token_generation}
\end{figure}

In order to generate a patch that replaces the original buggy code, we use \codebert to generate code token for every \masktoken in our input. To this end, we leverage the original training objective of mask replacement used in \codebert. 
\codebert is trained by predicting the correct token value that has been replaced with a \masktoken given its surrounding context. For each \masktoken, \codebert outputs a probability for each token in its vocabulary to replace the \masktoken. Typically, during training, a small percentage of tokens is masked out (< 15\%) and the model will attempt to predict the actual token that has been replaced. 

The task we have is similar to the original training objective of \codebert where we also preprocessed the inputs such that a small set of tokens has been masked out. However, \emph{a key difference between our input and the masked training data is our \masktoken{s} are grouped together}. A distinguished feature of \codebert and BERT family of models~\cite{devlin2019bert, liu2019roberta, feng2020codebert, guo2021graphcodebert} is the bidirectional nature where each token representation is predicated not only on context before but also context after. In order to generate replacement tokens for the \masktoken{s}, \codebert looks at the tokens before and after the mask location. This works well for the training objective since the \masktoken{s} are spread out where each token has sufficient tokens before/after to give context. However, for our input data, the \masktoken{s} are together. To generate an output for a \masktoken in the middle, the immediate context before/after are all \masktoken{s}. 

In order to facilitate token generation for grouped \masktoken{s}, we iteratively output tokens by replacing \masktoken{s} with previously generated tokens. Figure \ref{fig:token_generation}a shows an example of how this process is done. We start with the initial input \maskline of \CodeIn{if (endIndex <0 || <mask><mask><mask>) \{} and use \codebert to determine the top \beamwidth most probable replacement tokens for the first \masktoken. \beamwidth is the \textit{beam width} and is an adjustable parameter for our approach. In this example, \beamwidth is 3 meaning we take the top 3 most likely token along with its conditional probability. In the next iteration, we query \codebert again by replacing the first \masktoken with the top 3 replacement tokens (\CodeIn{1.startIndex, 2.endIndex, 3.emptyRange}) to find the top 3 token pairs with the highest joint conditional probability (\CodeIn{1.startIndex <, 2.startIndex ==, 3.endIndex >=}). We call this joint conditional probability value \textit{\tjs}. Given \(\tokenslength\) as the \masktoken length, \(\generatedseqfull\) as the \maskline with \(\tokensgeneratedlength\) tokens generated so far (\(\tokensgeneratedlength <= \tokenslength\)), let \(\codebertpredictortemp(\generatedseq, \generatedseqone)\) be the CodeBERT conditional probability of generating \(\generatedseqone\) when all tokens in \(\generatedseq\) after and including \(\generatedseqone\) have been masked out, the \tjs is defined as:
\begin{equation}
    \textit{\tjs(\generatedseq)} =\frac{1}{\tokensgeneratedlength} \sum_{i=1}^{\tokensgeneratedlength} \log(\codebertpredictortemp(\generatedseq, \generatedseqi))
\end{equation}
We note here that the \tjs does not represent the actual probability of the generated token once the complete line has been generated (all \masktoken{s} have been replaced). This is because \codebert uses both context before and after when determining the likelihood of a replacement token, the probability value does not account for the \masktoken{s} to be generated in future. When computing the \tjs of a token sequence (\(\generatedseqoneless\)) in the \maskline \(\generatedseq\), \codebert sees the values of the tokens before \(\generatedseqp\) (\(\generatedseqprev\)), however all tokens after are masked out (\(\generatedseqafter\)). That said, \emph{the \tjs is conditioned on the \masktoken{s} whose values have not yet been decided, and the conditional probability does not account for the future concrete values of those \masktoken{s}}. Thus, we use this probability value as a proxy only  temporarily and further re-assign the likelihood after (described in Section \ref{patch_reranking}) for more precise patch ranking. In each iteration, we use the \tjs to keep only the top 3 highest score generated token sequences. We repeat this process until we finish generating tokens for all \masktoken{s} in our input.

By generating tokens sequentially, we guarantee that when generating any \masktoken{s}, \codebert has at least one side of the immediate context. This helps with generating more syntactically correct candidate patches since \codebert can use the previous immediate context to inform what the best next tokens should be. This process is similar to beam search commonly used in code or natural language generation tasks~\cite{sutskever2014sequence}. One difference is that traditional code generation can accurately calculate the likelihood of a sequence as the average of the log conditional probability of the generated tokens. For our approach, the naive average is only an approximation of the likelihood since the probability outputs for tokens in the beginning of the \maskline do not include future generated tokens. To address this issue, we further re-rank each candidate patch by re-querying \codebert to obtain an accurate likelihood value as shown in Section~\ref{patch_reranking}. 

\subsection{Patch Re-Ranking}
\label{patch_reranking}

The re-ranking procedure makes use of the \codebert model again. The key idea is to provide an accurate score (i.e. likelihood) for each patch after it is \emph{fully} generated for more effective patch ranking. We start with the complete patch with all the generated tokens. We then mask out only one of the tokens and query \codebert to obtain the conditional probability of that token. We apply the same process for all other previous \masktoken locations and compute the \js which is an average of the individual token probabilities. Given \(\tokenslength\) generated tokens in a sequence: \(\fullgeneratedseq\), let \(\codebertpredictor(\generatedseq, \generatedseqone)\) be the \codebert conditional probability when masking out only token \(\generatedseqone\) in the sequence \(\generatedseq\), the \js is defined as:

\begin{equation}
    \textit{\js(\generatedseq)} = \frac{1}{\tokenslength} \sum_{i=1}^{\tokenslength} \log(C(\generatedseq, \generatedseqi))
\end{equation}

The \js can now be understood as the conditional probability of the generated sequence (i.e. given both contexts before and after, what is the likelihood of the generated patch according to \codebert?). This is done for all patches generated across all mask generation strategies (complete, partial, and template mask). We divide it by \(n\) to account for the token length difference since different \maskline{s} have different numbers of \masktoken{s}.%

Figure \ref{fig:token_generation}b shows an example of the re-ranking process. We use the 3 patches from the patch generation example and for each of them mask out the first token and obtain the probability value from \codebert. We repeat this process for the other two tokens and finally we end up with \js{}s for all 3 patches. We use the \js{}s to provide a ranking for each patch. By re-querying \codebert to obtain the \js we can provide more accurate patch ranking that allows for prioritization when only a subset of generated patches can be validated.

\subsection{Patch Validation}

For each candidate patch we generate, we apply the corresponding changes to the buggy file. We compile each patch and filter out any patches that fail to compile. We then run the test suite against each compiled patch to find plausible patches that pass all the tests. 

\label{patch_validation}

%% file: experiment.tex
\section{Experimental design}

\subsection{Research Questions}

In this paper, we study the following research questions:
\begin{itemize}
\item \textbf{RQ1:} How does \tech compare against state-of-the-art \apr tools?
\item \textbf{RQ2:} How do different configurations impact the performance of \tech?
\item \textbf{RQ3:} What is the generalizability of \tech for additional projects and multiple programming languages?
\end{itemize}

We demonstrate the effectiveness of \tech by comparing against both state-of-the-art traditional and \learning \apr tools with \textit{perfect fault localization} - the exact fix location of the bug is provided and \textit{not perfect fault localization} - use the suspicious locations generated by fault localization~\cite{wong2016fl} as inputs. 
Note that the former is the preferred or only comparison setting for all recent learning-based techniques since it eliminates the impact of other factors (such as fault localization) and can show the pure potential of different patch generation strategies~\cite{lutellier2020coconut, tufano2018empstudy, zhu2021recoder, jiang2021cure}. Therefore, this paper also uses perfect fault localization by default unless specifically mentioned.
We also show the contribution for each of our design components by conducting an ablation study. Finally, we evaluate the generalizability of \tech to additional projects in \dfj2.0 and QuixBugs. Additionally, we also evaluate multilingual repair capability of \tech by testing on the Python version of QuixBugs. 

\subsection{Implementation}\label{sec:impl}

\tech is implemented in Python with PyTorch~\cite{PyTorchWebPage} implementation of the \codebert model. We directly reuse the model parameters of the pre-trained \codebert model. For perfect fault localization patch generation, we use a beam width of 25 and generate at most 5,000 patches which is comparable to other baselines \cite{jiang2021cure, lutellier2020coconut}. For not perfect fault localization patch generation, we use a beam width of 5 and consider the top 40 most suspicious lines same as the recent \recoder tool~\cite{zhu2021recoder}.
We use Ochiai fault localization~\cite{abreu2007ochiai}, same as previous approaches~\cite{zhu2021recoder, liu2019tbar}. For patch validation, we use the UniAPR tool~\cite{chen2021uniapr}. All patches generated are validated and we evaluate \tech on an 8-core workstation with Intel i7 10700KF Comet Lake CPU @3.80GHz and 16GB RAM, running Ubuntu 20.04.3 LTS and OpenJDK Java 64-Bit Server version 1.8.0\_312 with NVIDIA GeForce RTX 3080 Ti GPU. For all our experiments, we set a time-out of 5-hour end-to-end time limit for fixing one bug, consistent with previous\learning tools~\cite{lutellier2020coconut, zhu2021recoder, li2020dlfix, saha2019hercules}. 

\subsection{Subject Systems}

For evaluation, we use the widely used benchmark of \dfj \cite{just2014dfj}. \dfj is a collection of reproducible bugs from open-source projects in Java. We first use \dfj version 1.2 to answer research questions 1 and 2. \dfj 1.2 contains 391 bugs (after removing 4 depreciated bugs) across 6 different Java projects. To address research question 3, we use \dfj 2.0 which adds 438 bugs on top of \dfj 1.2. Since \tech is designed for single line bug fixing, we evaluate only on the 82 single line bugs present in the new bugs in \dfj 2.0, this setup is similar to previous single line \apr tools~\cite{chen2018sequencer, jiang2021cure, lutellier2020coconut, zhu2021recoder}. We also use \quix~\cite{lin2017quixbug} that contains 40 small classic algorithms with single line bugs used to evaluate many \apr tools~\cite{lutellier2020coconut, jiang2021cure, zhu2021recoder, drain2021deepdebug, ye2019quixbugstudy, durieux2019estudy}. \quix contains both Python and Java versions of the same buggy programs. We evaluate \tech on both Python and Java versions to demonstrate the multilingual ability of our tool. 

\subsection{Compared Techniques}
We compare \tech against state-of-the-art baselines containing both \learning and also traditional \apr tools. For \learning \apr, we choose 6 recently published tools evaluated on Java or Python bug datasets: \recoder (Java) \cite{zhu2021recoder}, \deepdebug (Python)~\cite{drain2021deepdebug}, \cure (Java)~\cite{jiang2021cure}, \coconut (Java and Python)~\cite{lutellier2020coconut}, \dlfix (Java)~\cite{li2020dlfix}, and \sequencer (Java)~\cite{chen2018sequencer}. These tools use \nmt models to generate patches given the buggy line and surrounding context. Following the most recent \recoder work, we also compare against 12 state-of-the-art traditional single-hunk \apr tools: \tbar \cite{liu2019tbar}, \prapr \cite{ghanbari2019prapr}, \avatar \cite{liu2019avatar}, \simfix \cite{jiang2018simfix}, \fixminer \cite{koyuncu2020fixminder}, \capgen \cite{wen2018capgen}, \jaid \cite{chen2017jaid}, \sketchfix \cite{hua2018sketchfix}, \nopol \cite{demacro2014nopol}, \jgenprog \cite{martinez2015automatic}, \jmutrepair \cite{martinez2016astor} and \jkali \cite{martinez2016astor}. In total, our baseline comparisons comprise of 18 different \apr tools.

Following prior work~\cite{zhu2021recoder, jiang2021cure, lutellier2020coconut, liu2019tbar, ghanbari2019prapr},
we use patch correctness results gathered from previous papers \cite{zhu2021recoder, ghanbari2019prapr, jiang2021cure} for \dfj 1.2 evaluation and remove depreciated bugs. We use the recently updated results of \recoder by the authors~\cite{recoderUpdate} instead of the outdated results in the original paper~\cite{zhu2021recoder}. For many tools, we can only obtain either perfect fault localization or not perfect fault localization, therefore we only compare against baselines where the evaluation is under the same localization setting. For \dfj 2.0 evaluation, we directly run the two best performing baselines of \tbar (\template) and \recoder (\learning) 
with perfect fault localization under the same setting as our tool and report the results. For \quix evaluation, we compare against several \learning \apr tools since they have shown to perform the best on \quix for both Java and Python. All baseline results on \quix are taken from previous papers/experiments~\cite{zhu2021recoder, jiang2021cure, drain2021deepdebug}.

For evaluating our technique against state-of-the-art tools, we use the standard metrics of both \emph{plausible} patches that just pass the entire test suite of a project, and \emph{correct} patches that are syntactically or semantically equivalent to the developer patches. Following the common practice for \apr, the correct patches are determined by manually inspecting each plausible patch for semantic equivalency.

%% file: result.tex
\section{Result Analysis}

\subsection{RQ1: Comparison against state-of-the-art}

\begin{table*}
 \caption{Baseline comparisons with perfect fault localization}
 \centering
 \label{tab:comparison_soa}
 \scalebox{0.85}{
\begin{tabular}{|c||c|c|c|c|c|c|c|c|}
\hline
\textbf{Project} & \textbf{\tech} & \textbf{\recoder} & \textbf{\tbar} & \textbf{\cure} & \textbf{\coconut} & \textbf{\prapr} & \textbf{\dlfix} & \textbf{\sequencer}\\
 \hline
 \hline
Chart & 9 & 10 & 11 & 10 & 7 & 7 & 5 & 3\\
Closure & 23 & 21 & 16 & 14 & 9 & 12 & 11 & 3\\
Lang & 13 & 11 & 13 & 9 & 7 & 6 & 8 & 2\\
Math & 21 & 18 & 22 & 19 & 16 & 10 & 13 & 6\\
Mockito & 5 & 2 & 3 & 4 & 4 & 3 & 1 & 0\\
Time & 3 & 3 & 3 & 1 & 1 & 3 & 2 & 0\\
\hline
Total Correct / Plausible & \textbf{74 / 109} & 65 / 112 & 68 /
95 & 57 / 104 & 44 / 85 & 41 / 146 & 40 / 68 & 14 / 19\\
 \hline
\end{tabular}
}
\end{table*}

\begin{table}
 \caption{Baseline comparisons w/o perfect fault localization}
 \centering
 \label{tab:comparison_soa_wo_pfl}
 \scalebox{0.85}{
\begin{tabular}{|c||c|c||c|}
\hline
\textbf{Tool} & \textbf{Correct / Plaus.} & \textbf{Tool} &  \textbf{Correct / Plaus.} \\ 
 \hline
 \hline
\tech & \textbf{50 / 90} & \capgen & 22 / 25\\
\recoder & 49 / 96 & \jaid & 25 / 31\\
\avatar & 27 / 53 & \sketchfix & 19 / 26\\
\dlfix & 30 / 65 & \nopol & 5 / 35\\
\tbar & 42 / 81 & \jgenprog & 5 / 27\\
\prapr & 41 / 146 & \jmutrepair & 4 / 17\\
\simfix & 34 / 56 & \jkali & 1 / 22\\
\fixminer & 25 / 31 & &\\
 \hline
\end{tabular}
}
\end{table}

\begin{figure}
    \includegraphics[width=0.9\linewidth]{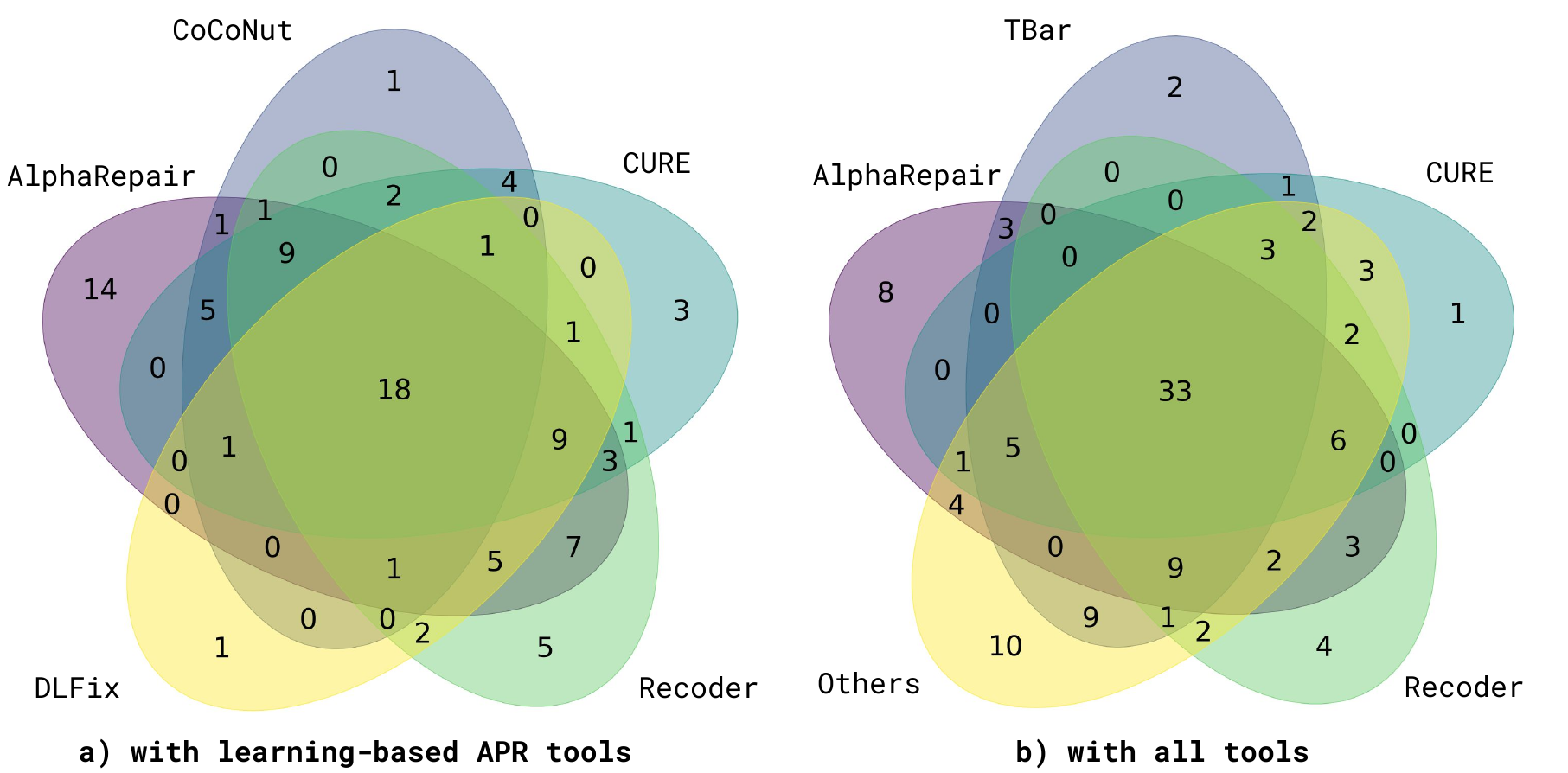}
    \caption{Correct patch Venn diagrams for \dfj 1.2}
    \label{fig:venn}
\end{figure}

\begin{figure}
    \includegraphics[width=\linewidth]{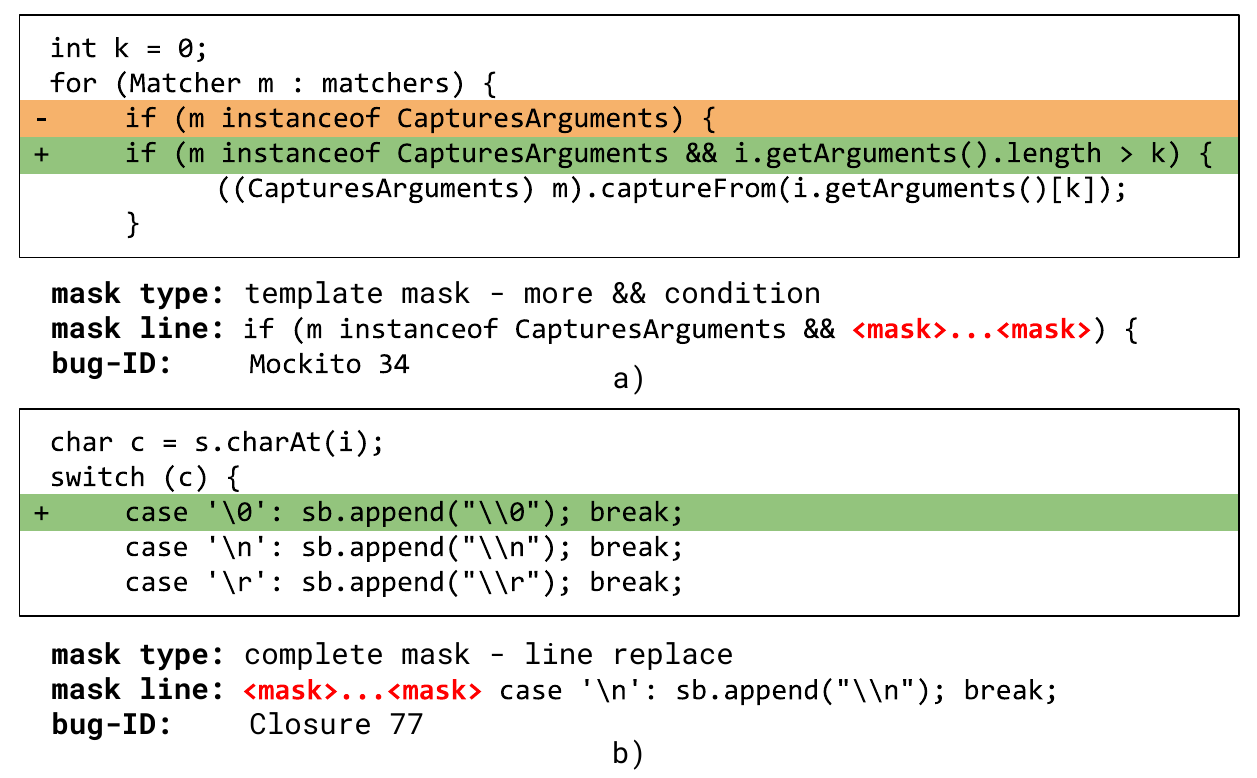}
    \caption{Example bug fixes in \dfj 1.2}
    \label{fig:unique_bugfix}
\end{figure}
\subsubsection{Perfect Fault Localization}
We first compare \tech with state-of-the-art \learning and traditional \apr tools under the preferred perfect fault localization setting. Table \ref{tab:comparison_soa} shows
the performance of \tech along with other baselines that also use perfect fault localization. \tech can successfully generate correct fixes for 74 bugs which outperforms all previous baselines including both traditional and \learning \apr techniques. 

To show the effectiveness of \tech further, we evaluate the number of unique bugs that only \tech can fix. We first compare against \learning \apr tools. Figure~\ref{fig:venn}a shows the unique fixes of \tech and other \learning tools (we exclude \sequencer since it has 0 unique bug fixes). We observe that \tech is able to fix the most number of unique bugs of 14. Figure~\ref{fig:venn}b shows the unique fixes of \tech, the 3 best performing baselines and all other \apr tools combined (Others in Figure~\ref{fig:venn}b). We observe that \tech is able to fix the most number of unique bugs of 8. This also demonstrates that \tech can be used together with other techniques to further increase the number of correct patches that can be generated. 

We provide a few examples of unique bugs that only \tech can fix. Figure~\ref{fig:unique_bugfix}a shows a bug with a missing length check on the array obtained from \CodeIn{i.getArguments()}. This is a difficult bug for both traditional and \learning tools to fix since the array used is not a variable but is obtained from a method call. Traditional \apr tools such as \template tools can detect that the method call returns an array, however it would be infeasible to add a length check for every such case as the search space would be too huge to traverse. \Learning tools rely on bug fixing changes for training data. While there could be many inserted length check fixes, this specific example, inserting a length check on a return value from a method call, can be rare to find in the dataset. \tech can fix this bug since the usage of the \codebert model does not require any bug fix code pairs and learns directly from large amount of open-source data where many of them contain similar code where the length checks can be placed on different expressions not just simple array variables. Furthermore, \tech also captures the context after and identifies the usage of \CodeIn{k} in accessing \CodeIn{i.getArguments()} to insert the correct length check. 

Figure~\ref{fig:unique_bugfix}b shows another bug that only \tech can fix. The correct fix is to insert an additional case statement to handle the missing case. This is a difficult bug to fix since it does not just slightly mutate any existing code line, but a completely new line needs to be added to handle a specific case in the program execution (when \CodeIn{c} is \CodeIn{\textbackslash0}). \tech can generate the correct fix for this bug by identifying its surrounding context. A case statement makes sense to insert here given the context of switch block and other case statements. \codebert is able to generate the appropriate case since other case statements use similar identifier formats (\CodeIn{\textbackslash n}, \CodeIn{\textbackslash r}). The outcome of the case also follows similarity to nearby context by adding block \CodeIn{sb.append(); break;}. Traditional \apr tools cannot fix this bug since it requires adding a new semantic line into the program which is beyond the ability of traditional \apr tools built for modifying existing lines or inserting simple statements (try catch, null pointer checker, etc). \Learning \apr tools also struggle with generating the correct patch for this bug since the added line does not fit a common edit pattern found in the training dataset. By observing the surrounding context and using previously seen examples (repeating case statements in other projects), \tech can generate the correct fix for this bug. These examples combined with the new state-of-the-art results achieved show that \tech opens up a new promising direction for \apr. 

\subsubsection{Not Perfect Fault Localization} We also compare against state-of-the-art tools without perfect fault localization. Table \ref{tab:comparison_soa_wo_pfl} shows the performance of \tech with other techniques also evaluated under this setting. \tech is able to produce 50 correct patches which outperforms previous state-of-the-art tools. Additionally, \tech is able to correctly fix \emph{7 unique bugs} (the highest among all studied techniques) that cannot be fixed by any other technique. For not perfect fault localization, since we do not have access to the ground truth location of the bug, \tech generates patches for multiple suspicious lines. To account for this, we lower the beam width of \tech for this setting in order to generate fewer patches per suspicious line. In this experiment, we show that even with the reduced number of patches generated per suspicious line, \tech can still achieve state-of-the-art results.

\subsection{RQ2: Ablation Study }

\begin{table}
 \caption{Component contribution}
 \centering
 \label{tab:component_contribut}
\scalebox{0.85}{
\begin{tabular}{|c||c|c|}
\hline
\textbf{Component} & \textbf{\#Correct Patch} & \textbf{\#Plausible Patch}\\
 \hline
 \hline
Complete mask & +20 & +29\\
Partial begin & +13 & +24\\
Partial end & +15 & +21\\
Template & +21 & +30\\
Comment buggy line & +5 & +5\\
 \hline
 Total & 74 & 109\\ 
 \hline
\end{tabular}
}
\end{table}

\begin{figure*}
    \includegraphics[width=0.8\linewidth]{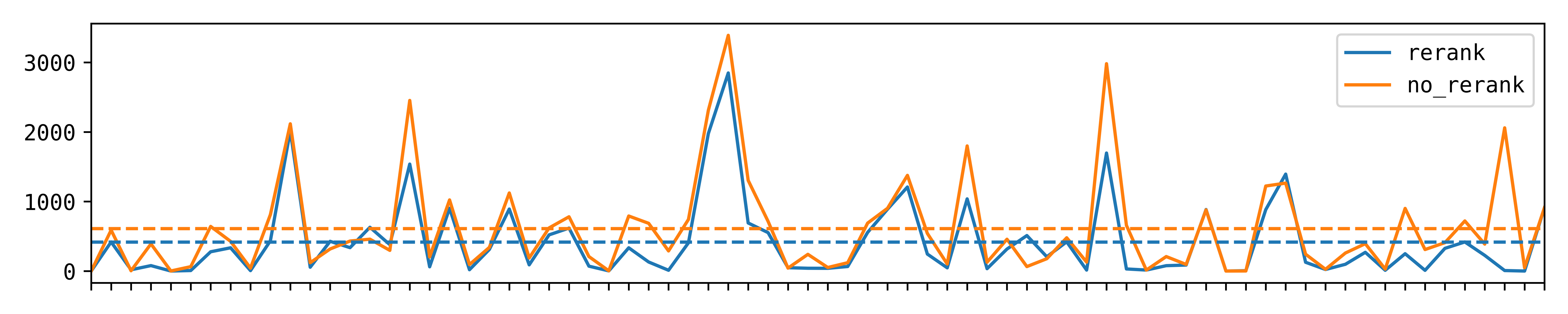}
    \caption{Patch ranking of 74 correct fixes with and without using patch re-ranking (lower is better)}
    \label{fig:rerank}
\end{figure*}

To study the contribution of adding different components in the design of \tech, we conduct
an ablation study. Table \ref{tab:component_contribut} contains the result with each 
row representing one component and the increase in number of correct/plausible patches \tech can produce. To show how each mask generation strategy (Section \ref{mask_generation}) improves the number of bugs fixed, we start with the most basic strategy and iteratively add more complex masking strategies. To begin with, we only use complete mask where the entire buggy line is replaced with all \masktoken{s}. This is the case where we give \codebert the entire freedom to generate any variety of edits. However, this is often not desirable as the search space grows exponentially with the number of \masktoken{s} and it becomes hard for \codebert to obtain a correct fix. We observe that we only achieve 20 correct patches when solely using this mask generation strategy. We obtain increases in correct patches generated as we start to use more mask generation strategies. The highest increase in performance is the usage of template \maskline{s} which add an additional 21 new fixes. Compared to complete mask, template mask only masks out certain parts of the buggy line (parameter, boolean expression, function calls). This allows \codebert to fill out only a small number of \masktoken{s} which limits the search space and allows \tech to quickly find the correct patch. In addition, we also see an increase in performance when we add the encoding for the commented version of the buggy line as input to \codebert. The buggy line itself is important for patch generation since it contains important information such as specific variables used and the type of line. This demonstrates that \tech is able to make use of the buggy line to help guide the generation of valid fixes. Combining all components in \tech we are able to achieve the final number of correct patches generated. 

 After the patch generation process, \tech re-queries \codebert again to generate more accurate ranking of each patch. To evaluate the effectiveness of our patch ranking strategy, we compare the order of the correct patches with and without re-ranking. Figure~\ref{fig:rerank} shows the patch ranking of all correct patches generated with and without re-ranking (dotted line represents the average patch ranking for each strategy). We observe that on average the correct patch is ranked 612th without using the re-ranking strategy. When using re-ranking, the correct patch on average is ranked 418th (31.7\% reduction). Furthermore, 61 out of 74 correct patches are ranked higher after re-ranking compared to before. As mentioned in Section \ref{patch_generation}, the \tjs (no re-ranking) is not an accurate representation of the actual likelihood of the generated tokens since it is conditioned on \masktoken{s} where the concrete values are not yet determined. By re-ranking the patches generated, we make sure that the \js is calculated without any \masktoken{s}, providing an accurate likelihood calculation. This demonstrates that the patch re-ranking process in \tech can effectively order the patches and prioritize patches that are ranked higher in case that only a subset of generated patches can be validated.

\subsection{RQ3: Generalizability of \tech}

\begin{table}
 \caption{Baseline comparisons on \dfj 2.0}
 \centering
 \label{tab:comparison_soa_d4j2}
 \scalebox{0.85}{
\begin{tabular}{|c||c|c|c|}
\hline
\textbf{Projects} & \textbf{\tech} & \textbf{\recoder} & \textbf{\tbar} \\
\hline 
\hline
Cli & 5 / 5 & 1 / 3 & 0 / 3\\
Codec & 6 / 7 & 2 / 4 & 1 / 3\\
Collections & 0 / 1 & 0 / 0 & 0 / 0\\
Compress & 1 / 3 & 1 / 3 & 1 / 2\\
Csv & 1 / 2 & 1 / 3 & 1 / 3\\
Gson & 2 / 3 & 0 / 1 & 0 / 0\\
JacksonCore & 3 / 3 & 2 / 3 & 1 / 2\\
JacksonDatabind & 8 / 9 & 2 / 2 & 1 / 4\\
JacksonXml & 0 / 0 & 0 / 0 & 0 / 0\\
Jsoup & 9 / 16 & 2 / 4 & 3 / 8\\
JxPath & 1 / 1 & 0 / 0 & 0 / 0\\
\hline
Total Correct / Plausible & \textbf{36 / 50} & 11 / 23 & 8 / 25\\
 \hline
\end{tabular}
}
\end{table}

\begin{figure}
    \includegraphics[width=\linewidth]{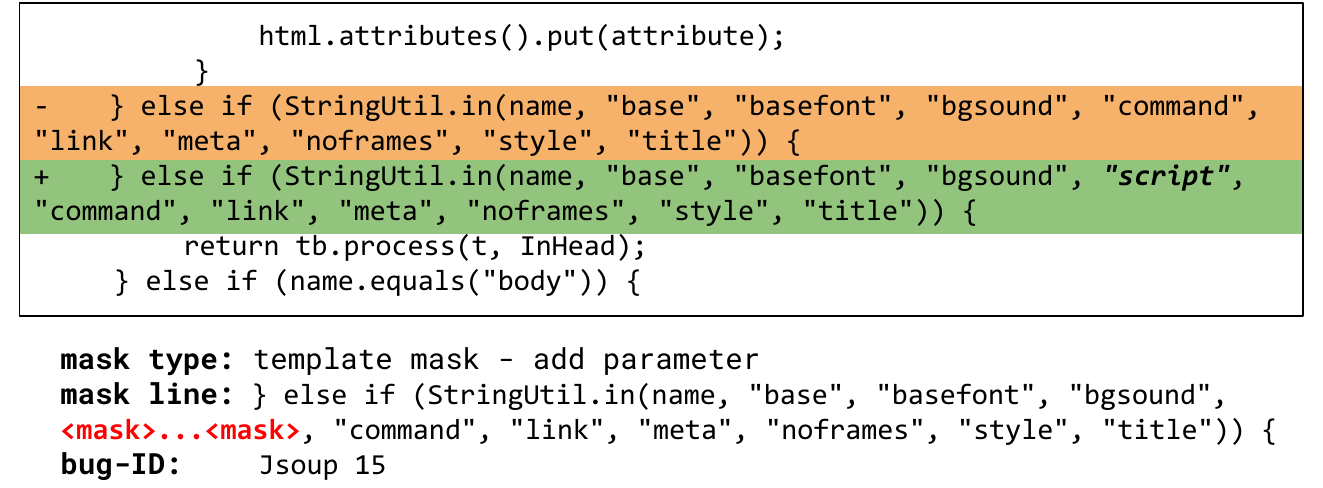}
    \caption{Example bug fix in \dfj 2.0}
    \label{fig:unique_bugfix_dfj2}
\end{figure}

\begin{table}
 \caption{Baseline comparisons on \quix}
 \small
 \label{tab:comparison_quixbug}
\scalebox{0.9}{
\begin{tabular}{|c||c|c|c|c|c|}
\hline
\textbf{Tool} & \tech & \cure & \deepdebug & \recoder & \coconut\\
\hline
\hline
\textbf{Java} & 28 / 30 & 26 / 35 & - / -& 17 / 17 & 13 / 20\\
\hline
\textbf{Python} & 27 / 32  & - / - & 21 / 22 & - / - & 19 / 21\\
 \hline
\end{tabular}
}
\end{table}

\subsubsection{\dfj 2.0}
To demonstrate the generalizability on additional projects and bugs and confirm that \tech is not simply overfitting to bugs in \dfj 1.2, we evaluate \tech on the 82 single line bugs in \dfj 2.0 dataset. Table \ref{tab:comparison_soa_d4j2} shows the results compared against other baselines on \dfj 2.0. We observe \tech is able to achieve the highest number of correct patches of 36 (3.3X more than top baseline). \dfj 2.0 contains a harder set of projects for \apr with different variety of fixes compare to \dfj 1.2. We observe that while \template tools such as \tbar was able to generate a high amount of correct patches for \dfj 1.2, the number of correct patches it can generate for \dfj 2.0 is limited. \Learning tools such as \recoder will also suffer from moving to a harder evaluation dataset since the edits are learnt from training datasets which might not be present in \dfj 2.0. In contrast, \tech does not use any fine-tuning on specific bug datasets which makes it less prone to suffer from  generalizability issues of traditional \template or \learning tools. 

Figure~\ref{fig:unique_bugfix_dfj2} shows an example of a bug from \dfj 2.0 dataset that only \tech can fix. In this example, the code checks if the variable \CodeIn{name} is one of the string literals. The bug is caused by missing a string literal of \CodeIn{"script"}. This bug is particularly hard to fix for both traditional and \learning \apr tools. For traditional tools such as \template ones, designing this specific pattern can be hard as it requires insertion of a seemingly arbitrary literal of \CodeIn{"script"}. For \learning tools, it faces the similar problem in lack of example bug fix pairs where the fix is to insert this particular string literal. In order to generate a correct fix of this bug, one must \textit{understand} the semantic meaning of the code. Upon further inspection, the string literals in this conditional statement are all HTML tags. \tech can generate the valid HTML string literal of \CodeIn{"script"} by understanding that the surrounding context deals with HTML documents and tags. Additionally, we observe other patches generated by \tech for this bug include other valid HTML tags such as \CodeIn{"head"}, \CodeIn{"html"}, \CodeIn{"font"}, etc. The specific example and improvement in repair effectiveness over the baselines demonstrate the generalizability of \tech. 

\subsubsection{\quix} We show the multilingual repair capability of \tech by evaluating on the \quix dataset, which contains both Java and Python versions of buggy programs. Table \ref{tab:comparison_quixbug} shows the results against state-of-the-art Java and Python \apr tools. We observe that \tech is able to achieve the highest number of correct patches in both Java and Python (28 and 27). We also observe that \tech is the only tool out of the baselines that can be directly used for multilingual repair (\coconut trains 2 separate models). Traditional \learning tools require access to bug fixing datasets which are often only in one programming language, restricting the ability for them to be used in a multilingual setting. Unlike traditional \learning \apr tools, \codebert is jointly trained on Java, Python, Go, PHP, JavaScript, and Ruby code snippets, this allows \tech to be directly used for multilingual repair tasks with minimal modifications. 

%% file: threats.tex
\section{Threats to Validity}

\mypara{Internal} One internal threat to validity comes from our manual analysis on the correctness of the patches. To this end, the authors carefully looked through all plausible patches and had detailed discussions in order to determine if a patch is correct. We have also released all correct patches for public evaluation along with the code to reproduce our experiments~\cite{correctPatchandDataset}.

Another internal threat is the direct usage of the \codebert model. The evaluation benchmark of \dfj could overlap with the training data used in \codebert which consists of over 6 million code functions. To address this, we calculated the number of fixed functions in \dfj that are in the \codebert training dataset. Overall, there are 65 out of 391 (16.6\%) \dfj 1.2 bugs and 9 out of 82 (11.0\%) \dfj 2.0 bugs that are present in the original training data. Out of the 74 and 36 bugs that \tech can correctly fix in \dfj 1.2 and 2.0, 10 and 5 (13.5\% and 13.9\%) bugs have their corresponding developer patch in the \codebert training data.
For the 15 bugs, we manually perturb the buggy code (change variable names, add empty while, if statements, etc) and use the perturbed version for repair. We observe that \tech is still able to generate the correct fixes for all 15 bugs. We believe this adequately shows that \tech is not simply overfitting to patches that are present in the original \codebert training dataset. Furthermore, the overall comparison results if we were to exclude the 15 overlapping bug fixes would still improve on state-of-the-art baselines (64 vs 63 on best baseline in \dfj 1.2 and 31 vs 10 on best baseline in \dfj 2.0). Note \quix dataset is not part of the \codebert training data. Future work to address this even more is to retrain the entire \codebert model by taking out all patched functions in original data and then re-evaluate \tech. 

Additionally, another internal threat is the experimental setup causing potential differences in results. For example, a longer timeout threshold or faster machine can lead to more bug fixes. To this end, we adopt an ordinary machine configuration (detailed in Section~\ref{sec:impl}) and follow prior \learning \apr tools~\cite{lutellier2020coconut, zhu2021recoder, li2020dlfix, saha2019hercules} by setting a 5-hour end-to-end timeout for fixing each bug. Furthermore, we follow the common practice in \apr by directly taking bug fix results from previous studies instead of directly running the \apr tools. To completely address this threat, one would need rerun the results from all the selected baselines \apr tools on the same machine with the same time-out threshold.

\mypara{External} The main external threat to validity comes from the evaluation benchmarks we chose. Our claims on the performance of \tech may not translate to other datasets. To address this threat, we evaluate the generalizability of \tech on a newer dataset - \dfj 2.0. We also evaluate our claim on the generalization to other programming languages by studying \tech on both the Python and Java versions of \quix. 

%% file: conclude.tex
\section{Conclusion}

We propose and implement \tech, the first \cloze \apr technique that leverages large pre-trained code model directly for repair under a zero-shot learning setting. This opens a new dimension for multilingual \learning \apr that does not require any fine-tuning on repair datasets. We build \tech using \codebert and design inputs to make use of the pre-training objective of \codebert to directly generate fix lines from the surrounding context. We evaluate \tech on popular Java benchmarks of \dfj and \quix to show that \tech achieves new state of the art with the highest improvement being 3.3X more bugs fixed than best baseline in \dfj 2.0. We further demonstrate the multilingual ability of \tech on the Python version of \quix where we achieved similar results compared to Java.